\documentstyle[prl,aps,multicol,epsf]{revtex}
\begin{document}
\draft
\title{\bf{Theory of Hysteresis Loop in Ferromagnets}}
\author{Igor F.Lyuksyutov$^{a,*}$, Thomas Nattermann $^{b}$,
and Valery Pokrovsky $^{a,c}$}
\address{(a) Department of Physics, Texas A\&M University,
College Station, TX 77843-4242 \\
(b) Institut f\"ur Theoretische Physik,
Universit\"at zu K\"oln, 50937, K\"oln, Germany\\
(c) Landau Institute for Theoretical Physics, Moscow, Russia}
\date{\today}
\maketitle

\begin{abstract}
A theory of the hysteresis loop in ferromagnets controlled by the
domain wall motion is presented. Domain walls are considered as
plane or linear interfaces moving in a random medium under the action
of the external ac magnetic field  $H=H_0\sin\omega t$. 
We introduce important characteristics
of the hysteresis loop, such as dynamic threshold fields, reversal field etc.
together with well known characteristics as coercive field and hysteresis
loop area (HLA) $\cal A$. We show that all these 
characteristics are regulated by two 
dimensionless combinations of the 
$H_0$ and $\omega$
and intrinsic characteristics of the ferromagnet. 
The moving domain wall can create
magnetic bubbles playing the role of pre-existing nuclei of the reversed
magnetization. We discuss a simple model of this process.
For magnetization reversal determined by  domain inflation 
we predict that HLA scales as ${\cal A}\propto \omega^{\beta}H_0^{\alpha}$ 
with $\alpha =1/2$ and $\beta =1/2$. Numerical simulations confirm 
this result.
\end{abstract}
\widetext
\begin{multicols}{2}

Magnetic hysteresis loops (HL) have been first studied already more
than a century ago  \cite{stein}. 
However, the understanding of this process 
in  thin magnetic films is still rather poor. 
Many efforts have been devoted recently to the prediction 
\cite{pelc,dhar,acharyya,zang} 
and experimental verification  \cite{wang1,wang2,jim}
of the scaling behavior of the {\it hysteresis loop area} (HLA)
as a function of the frequency and  
amplitude of the applied magnetic field 
in thin magnetic films \cite{jim}.
The scaling behavior of the HLA has been first reported 
in the pioneering work \cite{stein} for 3D magnets.
While there exists an extended literature on
the hysteresis of 3D magnets, the properties of the
HL in 2D systems are much less known
 \cite{wang1,wang2,jim,bruno,raquet}. 
Critical exponents found in the experiments with thin films 
differ dramatically for  different materials
 \cite{wang1,wang2,jim}
and probably for different regimes. Different authors
disagree with each other 
\cite{wang1,wang2,jim}
and also disagree with numerical
simulations \cite{pelc}.

More recently mean field type models with 
single \cite{bruno} or many 
\cite{raquet} relaxation times have been applied to 
analyze the experimental data for the description of the 
HL controlled by nucleation processes.
These authors predict a logarithmic dependence
of the coercive field $H_c$ on the  rate of the  applied
magnetic field $\dot H$.  
In a recent experiment \cite{jim} it was found  that  the HLA 
depends on frequency $\omega$ of the applied field 
as power with a small exponent $\alpha $
($\sim 0.03 - 0.06$) or, possibly, logarithmically.
In the framework of
the same approach the HLA
must behave also logarithmically in $H$. However, such a  dependence
has never been observed experimentally.

In this article we propose a new analysis of the  HL by 
formulating  a rather general
approach to the magnetization reversal mechanisms. 
In particular, we  indicate several important measurable 
characteristics of the HL besides of the HLA.
It turns out, that these characteristics 
are governed by two dimensionless parameters combined
from the field frequency $\omega$, its amplitude $H_0$
and characteristics of the magnetic material.
Everywhere in what follows we assume that the external
field varies harmonically in time, $H(t)=H_0\sin\omega t$.

In general, hysteresis behavior  may have various origins.
It can be mediated by nucleation processes,
by the domain wall (DW) propagation or simply by retardation of the 
magnetization due to fluctuations.
In this article we will restrict ourselves to the second mechanism,
i.e. the hysteretic behavior due to the finite 
velocity of the DW propagation,
and establish conditions at which it is dominant. 
Discussion of other mechanisms as well as a more detailed
presentation of the results of this paper will be postponed to a future 
publication \cite{LNP}. 

We consider magnets of Ising (uniaxial) symmetry.
Their properties may be very different depending of the
strength of the anisotropy. In the experimentally 
studied films the anisotropy was very weak. In this case
the domain wall width is large in comparison to the
lattice constant. On the contrary, in the original Ising
model the anisotropy is assumed to be large and DW width $l$
is simply the lattice constant.
However, these different models becomes equivalent after 
a simple rescaling:
the DW width should be accepted as a new elementary (cut-off)
length. It means that we consider a spin cluster of the linear size  $l$
as a new elementary spin. 

Disorder plays an important role in the DW propagation:
it creates a finite threshold value $H_p$ the external field has to
overcome to move the DW  and lowers the DW velocity considerably
in the low-field regime, $(H-H_p)\ll H_p$, where the DW motion shows
critical behavior \cite {nat92,fn93}. 
Peculiarities of the two-dimensional situation are:  much higher
mobility of the DW as well as much stronger fluctuations.
This makes the experimental situations much more diverse than those for 
a 3D magnet.

We first 
consider the individual DW motion by formulating an effective equation 
of motion under the influence of the field $H(t)$, which we solve then
in a finite geometry. 
This will lead us to the {\it characteristic fields}
$H_{t1}, H_{t2}, H_c, H_r$ and the HLA $\cal A$ 
which we analyze in several limiting regimes in which simple power
scaling is valid.

In general, the DW motion in an impure magnet is highly non-linear.
However, as it was shown in \cite{nat92,fn93}, 
after integrating out DW fluctuations on time scales less than the 
dynamical correlation time $t_v$ defined below,
the effective equation of motion for the 
center of mass coordinate of the interface $Z$ is given by
\begin{equation}
\dot Z\approx H_p\gamma f\left(\frac{H}{H_p}-1\right)\,,
\quad f(x)\approx\left\{\begin{array}{l@{\;,\;}l}
x^{\theta} & x\ll 1\\ x & x\gg 1\end{array}\right.
\label{dz/dt}
\end{equation} 
where  $\gamma$ represents the bare DW mobility.
Inside the critical region $H-H_p\ll H_p$,  $f(x)$ obeys a power law
with an exponent $\theta$ of the order $1/3$  
from an $\epsilon$-expansion around five dimensions extrapolated to 
to d=2 dimensions \cite{nat92}. 
On time scales $t\gg t_v$ (i.e. for $\omega t_v\ll 1$), and length scales 
$L \gg \xi_v$ 
where  
$t_v\approx \frac{l}{\gamma H_p}(\frac{H}{H_p}-1)^{-{\tilde \nu }}
\label{t_v}$ and $\xi_v \propto (\frac{H}{H_p}-1)^{-{\nu }}$,
fluctuations of the DW around its mean position are weak
and will be neglected therefore in the following. Here $\nu $ and ${\tilde \nu}$
are the correlation length and time, respectively,  exponent of the driven interface,
with $\nu (d=2)=1$ and ${\tilde \nu}(d=2)=4/3$
 again from an
$\epsilon$-expansion around five dimensions. Outside the critical region, 
$H\gg H_p$, $\theta=1$ and the domain wall is macroscopically flat.

We solve now equation of motion for a planar DW which can move
over a distance $L$ forth and back. Here $L$ corresponds to a
characteristic size of the system. 
In the simplest case $L$ denotes
its linear extension. In some experiments only one
domain wall survives and the model problem
of rectilinear domain wall motion is close to reality. Another
situation may occur in systems with rare, but strong extended
defects. In passing such a defect domain walls 
may form bubbles of reversed spins. These bubbles play then the role of
prepared nuclei in the next half-cycle of the magnetization
reversal. In this case $L$ can be identified with the mean distance $L_{ed}$
between the extended defects. E.g. in random field magnets strong 
defects result from a coherent fluctuation of random fields on 
neighboring sites in which case $L_{ed}$ 
is given by
\begin{equation}
L_{ed} \approx l \exp { \left [c \Gamma \over {(\eta - H_0)l}\right]^d}
\label{L_ed}
\end{equation}
where $\eta >H_0$ and $\Gamma$ denote the strength of the 
uncorrelated random field and the DW stiffness, respectively. Here 
$ \Gamma/l \gg \eta \gg H_p$ and c is
a constant of order unity.

The domain wall is assumed to be fixed at the left boundary of the
sample $Z\,=\,0$ at the initial moment.
Solving equation of motion (\ref {dz/dt}) for the domain wall 
coordinate $Z$
in harmonically oscillating magnetic field, we replace the integration
over time by the integration over the field, using 
$
dt=\frac{1}{\omega}\frac{dH}{\sqrt{H_0^2-H^2}}
$.
This  yields
\begin{equation}
Z(H)\,=\,{\gamma H_p\over\omega}\int_{H_p}^H f\left(\frac{H-H_p}{H_p}\right)
\frac{dH}{\sqrt{H_0^2-H^2}}.
\label{z(H)}
\end{equation} 
This equation is correct for $H>H_p$. 
At smaller values of $H$ the domain wall does not move.
The sign of
the square root changes each time as $H$ reaches its maximum or minimum
value $\pm H_0$.
The second necessary prescription is to substitute $H-H_p$ by
$-(H-H_p)$ in the argument of the function $f$ and $H_p$ by $-H_p$ in
the lower limit of the integral (\ref{z(H)}) when $H$ is negative.
To transfer from the coordinate $Z$ to the magnetic moment $\cal M$,
we write ${\cal M}(Z)\,=\,{\cal M}_s(2Z-L)/L$.
The area of hysteresis loop $\cal A$ can now be  expressed in  
integral form:
\begin{equation}
{\cal A}\,=\,\,4H_p{\cal M}_s(1-\frac{1}{LH_p}\int_{H_p}^{H_r}Z(H) \,dH\,).
\label{A1}
\end{equation}
For later applications, it is useful to introduce the functions
$\Phi(a,b)$ and ${\tilde \Phi}(a)$       
defined by 
\begin{equation} 
\Phi(a,b)  = \int_1^{a}f(x-1)\frac{dx}{\sqrt{b^2-x^2}}.
\label{Phi}
\end{equation}
and  $\tilde {\Phi}(a)=\Phi (a,a)$ . Eq. (\ref {z(H)}) can then 
be rewritten in the form $Z/L= u^{-1}\Phi (H/H_p,v)$, where 
$u$ and $v$ are the two dimensionless variables
\begin{equation}
u=\omega L/(\gamma H_p), \quad v=H_0/H_p
\label{uv}
\end{equation}
It is clear, that the HLA ${\cal A}$ also
depends on these two parameters.

There are two important characteristic values for the amplitude $H_0$
which separate hysteresis loops of different shapes. The first of them
is the {\it dynamic threshold field} $H_{t1}$, which is the smallest value of 
$H_0$ at which the domain wall reaches
the right boundary of the sample $Z=L$. At $H_0<H_{t1}$ the magnetization
is not reversed fully, the hysteresis loop is asymmetric, whereas at 
larger values of $H_0$ the hysteresis loop is 
symmetric under inversion 
$H\longrightarrow -H, {\cal M}\longrightarrow -{\cal M}$
(Fig. 1). The value of $H_{t1}$ is determined by 
solution of equation

\begin{equation} 
\frac{\omega L}{2\gamma H_p}  = \tilde {\Phi}(\frac {H_{t1}}{H_p})
\label{H_t1}
\end{equation}
At $H_0=H_{t1}$ the hysteresis loop is symmetric with respect
to reflections in the axis $H$ and $\cal M$.
 The second threshold field $H_{t2}$ is defined as a value of $H_0$
at which the domain wall reaches the right end of the sample $Z=L$
during one fourth of period, just at $H=H_0$. 
The definition of $H_{t2}$ differs from Eq. (\ref{H_t1}) by the 
absence of factor 2 in denominator of the l.h.s. 
The hysteresis loops corresponding to $H_0>H_{t2}$ acquire
characteristic ``mustaches'', single-valued pieces of the curve 
${\cal M}(H)$, which are absent in hysteresis curves for $H_0<H_{t2}$
(see Fig. 1d).

At a fixed $H_0>H_{t1}$ it is possible to define the {\it coercive field}
$H_c$ by the requirement ${\cal M}(H_c)=0$.
The so-called {\it reversal field} $H_r$ is defined as 
the field value at which the magnetic moment reverses fully.
At $H=H_r$ the two branches of the hysteresis curve intersect each other, 
i.e. $Z(H_r)=L$. At $H$ between $H_r$ and
$H_0$ the magnetic moment remains a constant ${\cal M}={\cal M}_s$.
The values $H_c$ and $H_r$ are shown in Fig. 1c,d.  
Using equation (\ref{z(H)}) and the relation for ${\cal M}(z)$, we
find for $H_c$:
\begin{equation}
\frac{\omega L}{2\gamma H_p}=\Phi(\frac{H_c}{H_p},\frac{H_0}{H_p}).
\label{H_c}
\end{equation}
For $ H_{t2}<H_0$ the relation for $H_r$ 
differs from (\ref{H_c}) by the absence 
of the the factor $1/2$ in the l.h.s. For $ H_{t1}<H_0<H_{t2}$ 
the reversal field $H_r$ is given by
\begin{equation}
\frac{\omega L}{\gamma H_p}=2{\tilde\Phi}(\frac{H_0}{H_p})
-\Phi(\frac{H_r}{H_p},\frac{H_0}{H_p}) 
\label{H_r}
\end{equation}
Thus, also the ratios $H_c/H_p$ and $H_r/H_p$ are functions of 
$u$ and $v$ only.

We analyze different asymptotics of the hysteresis loop
characteristics starting with the simplest limiting case $u\gg 1$. In this case 
the contribution to $Z(H)$ from the critical 
region is negligible and no further restriction for the 
applicability of (\ref {dz/dt}) apply. The
main contribution to the integral (\ref {Phi}) comes from large values of
the argument and the function $\Phi(a,b)$ can be calculated explicitly by 
replacing $f(x-1)$ by $x$ and setting the lower cut-off for the
integration equal to zero. 	This yields $\Phi(a,b)\approx b-\sqrt{b^2-a^2}$. 
As a result $H_p$ disappears from all expressions and a {\it new 
dimensionless parameter}
${\tilde u}=\omega L/(\gamma H_0)$
emerges. In particular
we get for dynamic threshold field $H_{t1} \gg H_p$ 
\begin{equation}
H_{t1}\,=\,\frac{\omega L}{2\gamma} = H_{t2}/2.
\label{h-t-3}
\end{equation}
Similarly, $H_c$ and $H_r$ are determined by 
$
H_c\,=\,\sqrt{H_{t1}(2H_0-H_{t1})}$ and $ H_r=\sqrt{H_{t2}(2H_0-H_{t2})}
$.
 The shape of the HL for $H_0>H_{t2}$ is described by
\begin{equation}
{\cal M}\,=\,{\cal M}_s\left[1\,-\,2{\tilde u}(1-\sqrt{1-(H/H_0)^2})\right].
\label{M}
\end{equation} 
It can be used also for the range of amplitudes
$H_{t1}<H_0<H_{t2}$ on the lower branch of the hysteresis curve $0<H<H_0$.
On the upper branch of this curve $H_r<H<H_0$ the sign of the
square root must be reversed.
If $H_0>H_{t2}$, the HLA is determined as follows:
\begin{equation}
\frac{{\cal A}}{4{\cal M}_sH_r}\approx 
1-{\tilde u}+\frac{{\tilde u}\arcsin\frac{H_r}{H_0}}
{2\sqrt{1-(\frac{H_r}{H_0})^2}}+{{\tilde u}\over2}\sqrt{1-(\frac{H_r}{H_0})^2}.
\label{A2}
\end{equation}
On the contrary, for $H_0<H_{t2}$, the area is:
\begin{equation}
{\cal A}\,=\,4{\cal M}_sH_r\left(1-{\tilde u}
\right)\,+\,\pi{\cal M}_sH_0{\tilde u}.
\label{A3}
\end{equation}
In the range of existence
of the full HL ${\tilde u}<1/2$. Therefore ${\tilde u}$ 
cannot be large. However it can be very small, i.e. $\omega L\ll \gamma H_0$.
In this case we find
$H_r\,\approx\,H_c\,=\,H_0\sqrt{\tilde u}=\sqrt{\omega L
H_0/\gamma}$,  
${\cal M}(H)\,=\,{\cal M}_s\left( 1\,-\,2H^2/({\tilde u}H_0^2)\right)$
and
\begin{equation}
{\cal A}\,\approx\,4{\cal M}_sH_r \propto\omega^{1/2}H_0^{1/2}.
\label{A4}
\end{equation}
Thus, for ${\tilde u}\ll 1$ we find the scaling behavior of all hysteresis
characteristics with universal critical exponents, independent on
the pinning centers. Therefore, one can expect that the same scaling
is valid for a clean ferromagnet with the relaxational dynamics.

Next we study the opposite limit of small $u\ll 1$.
In this case $H_{t1}$ and $H_{t2}$ are close to $H_p$ as follows from 
(\ref {H_t1}), (\ref {H_c}), (\ref {H_r}). Solving the 
integral (\ref {Phi}) approximately 
and employing
for $f(x)$ at small $x$ (\ref{dz/dt}), we get
\begin{equation}
\Phi(a,b) \approx \frac{(b-1)^{\theta+1/2}}{\sqrt 2} B(\theta+1,\frac{1}{2},\frac{a-1}{b-1})
\label{Phi2}
\end{equation}
where we introduced the incomplete $B$-function defined by the integral
\begin{equation}
B(x,y,z)=\int_0^z s^{x-1}(1-s)^{y-1}ds
\label {B}
\end{equation}
In particular, we get $\tilde \Phi(a) 
\approx \sqrt{\frac{\pi}{2}}(a-1)^{\theta+1/2}\frac{\Gamma(\Theta+1)}{\Gamma(\Theta+3/2)}
$. One then finds from (\ref {H_t1})  for $H_{t1}$
\begin{equation}
{H_{t1}\over H_p}-1\approx\left[\frac{(1+1/2\theta)\Gamma(\theta +1/2)}
{\sqrt{2\pi}\Gamma(\theta )}\frac {\omega L}{\gamma H_p}\right]^{{2\over 2\theta +1}}.
\label{h-t-2}
\end{equation}
$H_{t2}$ is given by the same expression if we replace $L$ by $2L$ in
(\ref{h-t-2}).
The ratio
$
(H_{t2}-H_p)/(H_{t1}-H_p)=2^{2/(2\theta+1)}
$
does not depend on $u$.
We omit here the explicit expressions for $H_c$ and $H_r$ but give finally
the results for the HLA in the case $u\ll 1$.
For (i) $v-1\ll 1$ we get:
\begin{equation}
\frac{{\cal A}}{4{\cal M}_sH_p}\approx 1+\frac{\sqrt{2}(v-1)^{\theta +1/2}}
{u}\int\limits_{0}^{w}\,\, B(\theta +1, 1/2; x)dx.
\label{A5}
\end{equation}
where $w=(H_r-H_p)/(H_0-H_p)$.
In the case (ii) $v\gg 1, uv\ll 1$ we find:
\begin{equation}
{\cal A}\,\approx\,4{\cal M}_sH_p\left[ 1\,+\,\frac{\left( (\theta +1)
uv\right)^{1/(\theta +1)}}{\theta +2}\right].
\label{A6}
\end{equation}
Finally, for (iii) $uv\gg 1$, $H_r\gg H_p$ and therefore all essential results 
formally coincide with those for $u\gg 1$ .
In the cases (i) and (ii) $\cal A$ is close to ${\cal A}_0=4{\cal
M}_sH_p$ 
but the deviation ${\cal A} - {\cal A}_0$ scales with the parameters
$u$ and $v$. The prerequisite inequalities  
$\omega t_v \ll 1$ and  $\xi_v \ll L$, 
lead  for $u \ll 1$ to further restrictions on  $L$ and $\omega$,   which
in general can be  fulfilled if $L$ is large enough {\cite {LNP}}.

We have also used  Monte-Carlo simulation 
with Glauber dynamics 
to check (\ref {dz/dt}),(\ref {M}) and (\ref {A4})
for a random bond Ising model.
The disorder has weak influence on phase 
diagram, but decreases significantly the DW velocity.
However, no measurable pinning threshold field $H_p $ 
has been detected.
To separate domain growth from domain 
nucleation we have studied two different
cases: (i) specially prepared defects for
fast nucleation, (ii) generated nucleus
with the opposite direction of  
magnetization in the form of circle or stripes.
The magnetization changed according 
to the model of a straight DW, i.e. similar 
to Eq. (\ref {M}).
In this case usually one or 
two domains inflate from the nucleation centers.
The distance $L_N$ between nucleation centers
play the role similar to the 
system size  $L$ in the above consideration.
We have studied scaling behavior of HLA $\cal A$ (see Eq. \ref {A4})
for the case when it is  determined by the domain growth. 
Details will be published
elsewhere \cite{LNP}.
We found scaling ${\cal A}\propto \omega^{\beta} $
with  $\beta = 0.49\pm 0.05 $ 
over  three decades
of $\omega $. However
the range of $H_0$ values available
is only one decade. For this reason we have
only checked that the dependence 
is consistent with  $ {\cal A} \propto H_0^{1/2} $.

To conclude, we have studied the hysteresis process in controlled by
the DW motion. We introduced two dynamical threshold fields
$H_{t1}$ and $H_{t2}$ corresponding to the occurrence of the full
magnetization reversal and to the occurrence of the single-valued
parts on the hysteresis curve respectively. These dynamical threshold fields 
are larger than the static threshold field $H_p$ which is required
to start the motion of the DW. We established that $H_{t1}$ and
$H_{t2}$ expressed in units $H_p$ are functions of one dimensionless
parameter $u=\omega L/\gamma H_p$ . The coercive
field $H_c$ and the reversal field $H_r$ expressed in the
same units are function of two
dimensionless parameters
 $H_c\,=\,H_pF\left(\frac{\omega L}{2\gamma H_p}, \frac{H_0}{H_p}\right)$ and
 $H_r\,=\,H_pF\left(\frac{\omega L}{\gamma H_p}, \frac{H_0}{H_p}
\right)$.
Experimental observation of this type of scaling would be the best
indirect evidence of the DW motion controlled hysteresis.
A direct observation of the DW motion in principle is possible
\cite{nikitenko}. At large fields 
$H_0\gg H_p$ the defects are inessential. Therefore the dependence
on $H_p$ must vanish from all scaling laws. It happens
indeed, and both fields $H_c$ and $H_r$ are expressed in terms of
one dimensionless parameter $\omega L/\gamma H_0$
We presented also corresponding
equations for the HLA to which the most experimental efforts were
concentrated. However, we would like to emphasize that the HLA is not
the only measurable characteristics of the HL and even not the
most informative of its characteristics: the fields $H_{t1}, H_{t2}, H_c,
H_r$ as well as the shape of the hysteresis curve are not less
interesting. 
In the case when the driven DW are almost free ($H_0\gg H_p$)
and the HL is narrow ($H_0\gg H_{t1}$) the HLA was found to be
proportional to $\omega^{1/2}H_0^{1/2}$. This conclusion is
supported by our numerical MC simulation.

Acknowledgments. This work was partly supported by the grants
DE-FG03-96ER45598, the SFB 341 and the GIF. V.P. is thankful to Prof.
J. Zittartz for the hospitality extended to him during his stay
at Cologne University.

\begin{figure}[b]
\begin{center}
\leavevmode
\epsfxsize=0.6\linewidth
\epsfbox{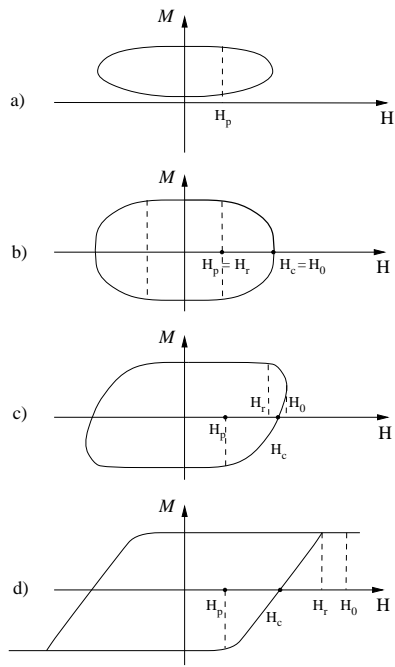}
\end{center}
\caption{ Schematic pictures of hysteresis loops.\\
(a) Incomplete HL for $H_0<H_{t1}$.\\
(b) Symmetric HL for $H_0=H_{t1}$.\\
(c) The HL for $H_{t1}<H_0<H_{t2}$.\\
(d) The HL for $H_0>H_{t2}$.\\
The values $H_p,\; H_c,\; H_r,\; H_0$ are marked in all figures.
}
\label{fig.dia}

\end{figure}

\end{multicols}
\end{document}